\documentclass[sigconf, nonacm, screen]{acmart}
\usepackage{times}  
\usepackage{array, graphics, graphicx, color, xspace, url, multirow, rotating, epsfig, amsmath, subfig}
\usepackage{wrapfig}
\usepackage{tikz}
\usepackage{booktabs}
\usepackage{epstopdf,subfig}
\usepackage{multicol}
\usepackage[format=plain,labelfont=bf,font=small]{caption}
\usepackage{ragged2e,flushend,enumitem,listings}
\usepackage{acronym} 
\usepackage{breakurl}
\usepackage{hyperref, cleveref}
\usepackage{multirow}
\usepackage[frozencache,cachedir=.]{minted}
\usepackage{appendix}
\usepackage[compact,small]{titlesec}
\usepackage[para]{footmisc}
\usepackage{comment}
\usepackage{paralist}

\hypersetup{pdfstartview=FitH,pdfpagelayout=SinglePage}

\newcommand\paragraphb[1]{\noindent{\bf{#1}}}
\newcommand\paragraphi[1]{\noindent\emph{#1}}
\newcommand\pb[1]{\paragraphb{#1}}
\renewcommand\pi[1]{\paragraphi{#1}}
\newcommand\pghi[1]{\paragraphi{#1}}

\newcommand{\bi}{\begin{itemize}}
\newcommand{\ei}{\end{itemize}}

\newcommand{\ie}{\emph{i.e.,}\xspace}
\newcommand{\eg}{\emph{e.g.,}\xspace}

\newcommand{\eat}[1]{}

\newcommand{\allnotes}[1]{}
\renewcommand{\allnotes}[1]{\textit{#1}} %

\hypersetup{                    %
  colorlinks,
  linkcolor={green!80!black},
  citecolor={red!70!black},
  urlcolor={blue!70!black}
}

\let\svthefootnote\thefootnote
\newcommand\freefootnote[1]{%
  \let\thefootnote\relax%
  \footnotetext{#1}%
  \let\thefootnote\svthefootnote%
}

\newcommand{\squishlist}{
  \begin{list}{$\bullet$}{
    \setlength{\itemsep}{0pt}       \setlength{\parsep}{3pt}
    \setlength{\topsep}{3pt}        \setlength{\partopsep}{0pt}
    \setlength{\leftmargin}{1em}    \setlength{\labelwidth}{1em}
    \setlength{\labelsep}{0.5em} } }

\newcommand{\squishend}{
  \end{list} }
\newcommand{\squishlistend}{
  \end{list} }

\graphicspath{{./dia/}}

\newcommand\vldbpagestyle{plain} 

\begin{document}
\title{Liberal Entity Matching as a Compound AI Toolchain (Extended Abstract)}

\author{Silvery D. Fu$^{1,2}$, David Wang$^{1}$, Wen Zhang$^{1}$, Kathleen Ge$^{1}$}
\affiliation{$^{1}$UC Berkeley, $^{2}$System Design Studio}

\begin{abstract}

As developers increasingly embrace the capabilities of new large language models (LLMs), the focus on AI application development is shifting from only focusing on models to developing compound systems with multiple components to achieve state-of-the-art results~\cite{compound-ai-blog}. In this paper, we explore the task of \emph{entity matching} (EM) with a compound AI system approach. EM is a fundamental problem in data management and integration, which involves determining whether two descriptions refer to the same real-world entity~\cite{vldb21-ditto}. For instance, consider determining if product descriptions on Ebay and Amazon (see Fig.\ref{fig:libem}) refer to the same product.

Entity matching has evolved from rule-based and distance-based approaches~\cite{vldb21-em-similarity}, to machine learning~\cite{cidr13-tamer,icde21-autoem}, crowdsourcing~\cite{vldb12-crowder, cidr13-tamer}, deep learning~\cite{sigmod18-deep-em}, and pre-trained language models (PLMs)~\cite{vldb21-ditto,vldb22-fmwrangle}, showing a continuing trend of improved results. Recently, the large language models (LLMs) based approach came to the fore and showed state-of-the-art results over multiple datasets~\cite{arxiv23-matchgpt, arxiv24-matchgpt, arxiv23-batcher} with prompt engineering such as hand-crafted rules and in-context learning (ICL). We refer to this as \emph{solo-AI EM}, which prompts LLMs to perform entity matching in a single model call. 

However, important challenges remain with the current solo-AI approach: (1) Solo-AI EM often relies on \textbf{hand-tuned} rules and examples as part of prompt engineering. To do this, humans often need to perform trial and error to find the right prompt for each new dataset. For example, an LLM may not know that under a given EM setting, color distinguishes different products and thus requires manual specification or hand-picked few-shot examples for the specific dataset. (2) Solo-AI EM relies on \textbf{static} knowledge from training data and thus may fail at entity matching for entities that appear beyond that time. For example, when new products have just been released, LLMs may fail to correctly recognize them for matching. (3) Entity matching is typically a stage in a larger data processing pipeline (\eg entity resolution) that involves other forms of data processing. Current solo-AI EM follows a \textbf{rigid} preprocessing of input data with predefined rules and does not allow for adapting the input data in a format best suited for the model and task. For example, as we will show, for several datasets, simply including schema information in the entity data---which is typically stripped off in existing systems---can improve the matching accuracy.

Further, the solo-AI approach makes it hard to iterate over the system designs. Existing solo-AI EM systems are usually in the form of Python notebooks~\cite{arxiv24-matchgpt}, as opposed to libraries that one can easily incorporate in their applications or APIs to use as a service. Such a lack of system modularity also makes it difficult to configure and optimize the EM system to navigate the trade-offs between accuracy and other performance metrics. 

\begin{figure}
    \centering
    \includegraphics[scale=0.34]{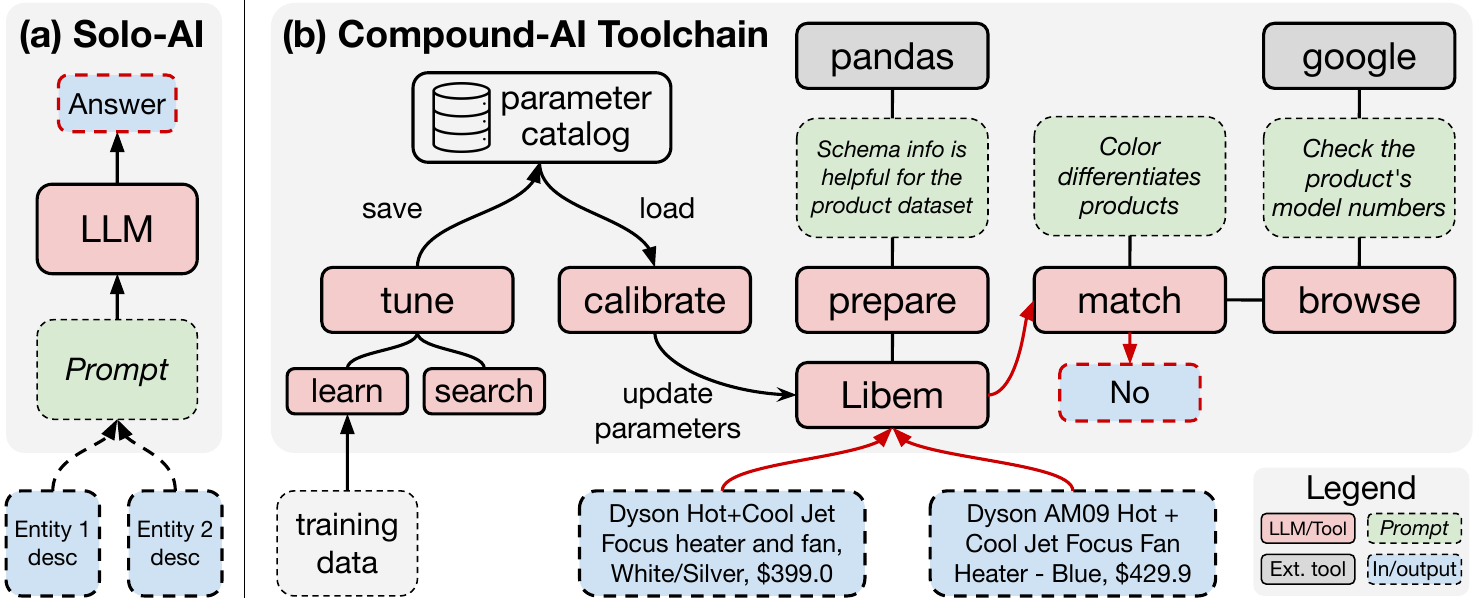}
    \caption{Entity Matching with Solo-AI vs. Compound-AI Toolchain. In Libem: (1) EM is performed liberally with LLM using tools such as data preprocessing (``prepare'' tool) and browsing external data sources (``browse'' tool); (2) Parameters to configure each tool can be learned when the training data and/or performance metrics are provided.}
    \label{fig:libem}
    \vspace{-0.2in}
\end{figure}

In this paper, we argue that entity matching should be performed \emph{liberally} by AI as opposed to being constrained by (1)-(3), to maximize its accuracy, performance, and ease-of-use. Our key insight is that given that LLMs can not only provide knowledge, but also behaviors such as invoking external tools, an EM system should provide proper \emph{tools} for LLMs so that they can solve the tasks better and even self-improve their performance. To achieve liberal EM, we argue that an EM system should be best designed as a compound AI system~\cite{compound-ai-blog} that consists of both AI and system components. Specifically, we want an EM \emph{toolchain} that provides:

\squishlist
    \item \textbf{Tool use.} The toolchain should provide relevant tools, such as data processing and information retrieval, so that a model can liberally decide what, when, and how to leverage tools to better perform the EM task.
    
    \item \textbf{Self-refinement}. The toolchain should adapt to the input dataset and improve its performance without hand tuning when training data is available. Inspired by DSPy~\cite{arxiv23-dspy, arxiv23-dspy-assertion}, we aim for the toolchain to start with simple, general prompts and evolve towards better, more task/dataset specific prompts and parameters that achieve higher matching accuracy.
    
    \item \textbf{Optimization.} Users should be able to easily configure and optimize (\eg turn off chain of thought in the browsing step to avoid lengthy reactions for search results) to navigate the trade-offs between performance and cost. Similar to self-refinement, the toolchain should be capable of automatic optimization.
\squishend

We propose \textbf{Libem}, a \emph{compound AI toolchain} that aims to perform entity matching through tool use, self-refinement, and optimization. To achieve this, we made several key design choices in Libem. First, instead of a collection of prompts, we structure Libem as a collection of composable and reusable modules/tools, as shown in Fig.\ref{fig:libem}, where each tool can also be individually invoked for testing or external (re)use. Second, we separate the parameters (\eg discrete configurations and prompts) from the organization of tools, allowing the parameters to be tuned and configured, as inspired by DSPy~\cite{arxiv23-dspy}. Each Libem tool has its own parameters and may invoke model calls, other Libem tools, or external APIs. For example, the match tool can take an entity pair and return a prediction, and may invoke the browsing tool, while the browsing tool can use Google Search, and the preparation tool can utilize Pandas. Libem can self-refine through training data and can save the optimal parameters for reuse. Third, upon each run of the Libem match, a calibration process takes into account the input dataset and the performance goals to configure Libem (\eg using the saved parameters) to best perform the incoming EM task.

\pb{Design overview.} Fig.\ref{fig:libem} describes the internals and typical workflow of entity matching with Libem. Each tool module has functions, interfaces, parameters, and prompts. The tools are organized in a hierarchy, where the parent provides the interface to access the tool. For example, the \emph{libem.interface} contains match, prepare, and tune tools, and the \emph{libem.match.interface} contains the browse tool.

\pghi{(1) EM with tool use.} When Libem attempts to perform a match, multiple model calls and tool calls may occur. Upon invoking Libem, \texttt{libem.match} is activated, and the model decides which tools to use given the tool's parameters. For instance, if the browse tool is selected, Libem will first search the web for the entity information and then use the retrieved data to perform the matching.

\pghi{(2) Self-refinement.} Libem supports self-refinement in the following manner: The \texttt{tune} tool invokes \texttt{libem.match} on training samples and obtains \emph{rules} or \emph{experiences} (\ie mistakes to avoid). We implement three simple forms of self-refinement strategies when training data is available: (a) generate rules from successful matches (\eg ``color differentiates entities''); (b) generate experiences from failed matches; (c) search for optimal parameters. Learned rules, experiences, and parameters are saved in a persistent catalog. When input entities are provided, the \texttt{libem.calibrate} tool determines which set of parameters to use with the match step. For example, if the input entities are products, Libem loads and calibrates its tools with corresponding parameters from the parameter catalog.

\pghi{(3) Extending the toolchain.} Additional tools can be easily added by defining new tools and adding them to the Libem toolchain. We include a simple code generator with the Libem CLI to facilitate the generation of boilerplate code for new tools.

\pb{Libem prototype.} Libem is under active development. Our current prototype consists of 2,589 SLOC in Python and includes the following top-level tools: \texttt{match}, \texttt{browse}, \texttt{prepare}, \texttt{tune}, \texttt{calibrate}, \texttt{optimize}, and sub-level tools such as \texttt{learn}, \texttt{search} in \texttt{libem.tune} (Fig.\ref{fig:libem}). Besides, we have added logging, tracing, and telemetry code to enable users to easily track and debug the system's behavior. Libem can be imported as a Python library, invoked via command line, or interacted as a web service.

\pb{Early experiments.} We evaluate Libem by running entity matching on real-world datasets covering product information and bibliographical data from Abt-Buy, Walmart-Amazon, Amazon-Google, DBLP-Scholar, and DBLP-ACM~\cite{vldb21-ditto, arxiv23-matchgpt}, as used in prior research~\cite{arxiv24-matchgpt, arxiv23-batcher}. We compare Libem to a solo-AI baseline and report precision, recall, and F1 score. We use GPT-4-turbo to process model calls.

\begin{table}[]
    \centering
    \footnotesize
    \begin{tabular}{|l|cc|cc|cc|}
    \hline
    \multirow{2}{*}{\textbf{Dataset}} & \multicolumn{2}{c|}{\textbf{Precision}} & \multicolumn{2}{c|}{\textbf{Recall}} & \multicolumn{2}{c|}{\textbf{F1}}\\
    \cline{2-7}
    & \textbf{S} & \textbf{C} & \textbf{S} & \textbf{C} & \textbf{S} & \textbf{C} \\ \hline
        Abt-Buy & 95.1 & 96.6 & 95.1 & 96.1 & 95.1 & \textbf{96.4} \\
        Walmart-Amazon & 84.3 & 94.05 & 94.3 & 81.87 & 89.0 & 87.53 \\
        Amazon-Google & 63.8 & 71.1 & 92.7 & 98.7 & 75.6 & \textbf{82.6} \\
        DBLP-Scholar & 89.7 & 90.6 & 87.2 & 96.0 & 88.4 & \textbf{93.2} \\
        DBLP-ACM & 94.0 & 96.5 & 100.0 & 99.6 & 96.9 & \textbf{98.0} \\ \hline
    \end{tabular}
    \caption{EM Accuracy with Solo- (S) vs. Compound-AI in Libem (C).}
    \vspace{-0.4in}
    \label{tab:f1}
\end{table}

Table~\ref{tab:f1} presents our initial findings. First, we want to point out that these datasets were released before the model was trained, thus our results run the risk of data leakage~\cite{balloccu2024leak}, as was also the case in prior work~\cite{arxiv23-batcher, arxiv24-matchgpt}. Therefore, we are collecting new datasets that were not included in model training, \eg from Bandai online shops and Bandai Wiki, featuring the newest releases in April 2024, also to test out the capabilities of the browsing tool. Nonetheless, as shown in Table~\ref{tab:f1}, Libem outperforms the solo-AI counterpart in four out of five existing datasets. We observe a 3\% increase in the average F1 score across the five datasets, with a maximum of 7\% in the Amazon-Google dataset. Due to space limitations, we summarize the following early findings from experimenting with Libem: (1) We can achieve better or comparable performance without manually tuning the prompts, compared to those that involve human-in-the-loop tuning, where self-refinement helps avoid common mistake patterns. (2) A simple choice of enabling schema helps substantially in matching accuracy. (3) Browsing can be highly beneficial when the data sources are recent; however, it often needs to be accompanied by explicit 'chain of thought' prompting to perform well.

\pb{Ongoing and future work.} We are actively developing Libem with a focus on the following fronts: (i) Better tooling. We are extending and enhancing the tools in the Libem toolchain, such as browsing on user-supplied data sources. (ii) Matching speed and efficiency. Optimizing performance metrics is often as crucial as accuracy. We are working on enhancing per-match latency, throughput (\eg with batching~\cite{arxiv23-batcher}), and token efficiency (\eg with caching~\cite{gptcache}). We are investigating mechanisms for dynamic trade-offs and optimizations between token usage, accuracy, and speed, such as generating classifiers like random forests from rules learned by Libem to be used in place of model calls during matching. (iii) Refinement strategies. We are exploring alternative strategies and algorithms for self-refinement and calibration, for both prompts and other parameters, including search algorithms such as Bayesian optimization and synthetic data generation~\cite{arxiv23-dspy}. (iv) Practicality. We are investigating how to better deploy and serve the toolchain efficiently, and how to measure and ensure its robustness, \eg with tracing and new programming primitives like assertions~\cite{arxiv23-dspy-assertion}. We plan to conduct large-scale evaluations with more datasets~\cite{arxiv23-batcher} and with open-source models~\cite{arxiv23-llama, dbrx-blog}.

Finally, we plan to apply the compound AI toolchain approach to broader tasks, such as entity resolution, data cleaning~\cite{arxiv24-seed}, and schema matching and mapping~\cite{icde23-plm-schema-match}, to develop practical compound AI solutions. The Libem library, examples, and benchmarks will be available as an open-source project at \url{https://libem.org}.

\end{abstract}

\maketitle

\pagestyle{\vldbpagestyle}

\vspace{-0.008in}
\bibliographystyle{ACM-Reference-Format}
\bibliography{libem}

\end{document}